\documentclass[12pt]{article}
\usepackage{amsmath}
\usepackage{amssymb}

\input{tcilatex}
\begin{document}

\bigskip

\bigskip \ 

\begin{center}
\textbf{SOME MATHEMATICAL AND PHYSICAL REMARKS}

\smallskip \ 

\textbf{ON SURREAL NUMBERS}

\bigskip \ 

\smallskip \ 

J. A. Nieto\footnote[1]{%
niet@uas.edu.mx,janieto1@asu.edu}

\smallskip \ 

\textit{Facultad de Ciencias F\'{\i}sico-Matem\'{a}ticas de la Universidad
Aut\'{o}noma}

\textit{de Sinaloa, 80010, Culiac\'{a}n Sinaloa, M\'{e}xico}

\bigskip \ 

\bigskip \ 

\textbf{Abstract}
\end{center}

We make a number of observations on Conway surreal number theory which may
be useful, for further developments, in both in mathematics and theoretical
physics. In particular, we argue that the concepts of surreal numbers and
matroids can be linked. Moreover, we established a relation between the
Gonshor approach on surreal numbers and tensors. We also comment about the
possibility to connect surreal numbers with supesymmetry. In addition, we
comment about possible relation between surreal \smallskip \ numbers and
fractal theory. Finally, we argue that the surreal structure may provide a
different mathematical tools in the understanding of singularities in both
high energy physics and gravitation.

\bigskip \ 

\bigskip \ 

\bigskip \ 

\bigskip \ 


Pacs numbers: 04.60.-m, 04.65.+e, 11.15.-q, 11.30.Ly

November, 2016

\newpage \noindent \textbf{1. Introduction}

\smallskip \ 

Surreal numbers is a fascinating subject in mathematics. Such numbers were
invented, or discovered, by the mathematician John Horton Conway in the 70's
[1]-[2]. Roughly speaking, the key Conway%
\'{}%
s idea is to consider a surreal number in terms of previously created dual
sets $X_{L}$ and $X_{R}$. Here, $L$ stands for left and $R$ for right. One
of the interesting things is that such numbers contains many well known
ordered fields, including integer numbers, the dyadic rationals, the real
numbers and hyperreals, among other numerical structures. Moreover, the
structure of surreal numbers leads to a system where we can consider the
concept of infinite number as naturally and consistently as any `ordinary'
numbers.

It turns out that in contrast to the inductive Conway definition of surreal
numbers, Gonshor [3] proposed in 1986 another definition which is based on a
sequence of dual pluses and minuses \{$+,-$\}. Gonshor itself proves that
his definition of surreal numbers is equivalent to the Conway definition.

In this article, we shall make a number of remarks on surreal number theory
which we believe can be useful in both scenarios: mathematics and physics.
In particular, we shall established a connection between surreal numbers and
tensors. Secondly, we shall show that surreal numbers can be linked to
matroids. Moreover, we shall argue that surreal numbers may be connected
with spin structures and therefore may provide an interesting developments
in supersymmetry. We also comment about the possibility that surreal numbers
are connected with fractal theory. Finally, we also mention that concepts of
infinitely small and  infinitely large in surreal numbers may provide a
possible solution for singularities in both high energy physics and
gravitation.

Technically, this work is organized as follows. In section 2, we briefly
review the Conway definition of surreal numbers. In section 3, we also
briefly review the Gonshor definition of a surreal number. In section 4, we
established a connection between surreal numbers and tensors. In section 5,
we comment about the possibility that surreal numbers and matroids are
related. Moreover, in section 6 we mention number of possible applications
of the surreal number theory, division algebras, supersymmetry, black holes
and cosmology.

\smallskip \ 

\smallskip \ 

\noindent \textbf{2. Conway formalism}

\smallskip \ 

Let us write a surreal number by

\begin{equation}
x=\{X_{L}\mid X_{R}\}  \label{1}
\end{equation}%
and call $X_{L}$ and $X_{R}$ the left and right sets of $x$, respectively.
Conway develops the surreal numbers structure $\mathcal{S}$ from two axioms:

\smallskip \ 

\textbf{Axiom 1}. Every surreal number corresponds to two sets $X_{L}$ and $%
X_{R}$ of previously created numbers, such that no member of the left set $%
x_{L}\in X_{L}$ is greater or equal to any member $x_{R}$ of the right set $%
X_{R}$.

\smallskip \ 

Let us denote by the symbol $\ngeq $ the notion of no greater or equal to.
So the axiom establishes that if $x$ is a surreal number then for each $%
x_{L}\in X_{L}$ and $x_{R}\in X_{R}$ one has $x_{L}\ngeq x_{R}$. This is
denoted by $X_{L}\ngeq X_{R}$.

\smallskip \ 

\textbf{Axiom 2}. One number $x=\{X_{L}\mid X_{R}\}$ is less than or equal
to another number $y=\{Y_{L}\mid Y_{R}\}$ if and only the two conditions $%
X_{L}\ngeq y$ and $x\ngeq Y_{R}$ are satisfied.

\smallskip \ 

This can be simplified by saying that $x\leq y$ if and only if $X_{L}\ngeq y$
and $x\ngeq Y_{R}$.

\smallskip \ 

Observe that Conway definition relies in an inductive method; before a
surreal number $x$ is introduced one needs to know the two sets $X_{L}$ and $%
X_{R}$ of surreal numbers. Thus, since each surreal number $x$ corresponds
to two sets $X_{L}$ and $X_{R}$ of previous numbers then one wonders what do
one starts on the zeroth day or $0$-day? If one denotes the empty set by $%
\emptyset $ then one defines the zero as

\begin{equation}
0=\{ \emptyset \mid \emptyset \}.  \label{2}
\end{equation}%
Using this, one finds that in the first day or $1$-day one gets the numbers

\begin{equation}
-1=\{ \emptyset \mid 0\},\text{ \  \  \  \  \  \  \  \  \  \  \ }\{0\mid \emptyset
\}=+1.  \label{3}
\end{equation}%
In the $2$-day one has

\begin{equation}
-2=\{ \emptyset \mid 1\},\text{ \ }-\frac{1}{2}=\{-1\mid 0\},\text{\  \  \ }%
\{0\mid 1\}=\frac{1}{2},\text{\  \  \  \ }\{1\mid \emptyset \}=+2.  \label{4}
\end{equation}%
While in the $3-$day one obtains

\begin{equation}
\begin{array}{c}
-3=\{ \emptyset \mid 2\},\text{ \ }-\frac{3}{2}=\{-1\mid 0\} \text{\  \  \ }
\\ 
\\ 
\frac{1}{2}=\{0\mid 1\},\text{\  \  \  \  \  \ }+2=\{2\mid \emptyset \}.%
\end{array}
\label{5}
\end{equation}%
The process continue as the following theorem establishes.:

\smallskip \ 

\textbf{Theorem 1}. Suppose that the different numbers at the end of $n$-day
are

\smallskip \ 

\begin{equation}
x_{1}<x_{2}<...<x_{m}.  \label{6}
\end{equation}%
Then the only new numbers that will be created on the $(n+1)$-day are

\begin{equation}
\{ \emptyset \mid x_{1}\},\{x_{1}\mid x_{2}\}...\{x_{m-1}\mid
x_{m}\},\{x_{m}\mid \emptyset \}.  \label{7}
\end{equation}%
Furthermore, for positive numbers one has

\begin{equation}
\{x_{m}\mid \emptyset \}=x_{m}+1  \label{8}
\end{equation}%
and

\begin{equation}
\{x_{m}\mid x_{m+1}\}=\frac{x_{m}+x_{m+1}}{2}.  \label{9}
\end{equation}%
While defining 
\begin{equation}
-x=\{-X_{R}\mid -X_{L}\},  \label{10}
\end{equation}%
for negative numbers one gets

\begin{equation}
\{ \emptyset \mid x_{m}\}=-(x_{m}+1)  \label{11}
\end{equation}%
and

\begin{equation}
\{-x_{m}\mid -x_{m+1}\}=-\frac{(x_{m}+x_{m+1})}{2}.  \label{12}
\end{equation}%
Thus, at the $n$-day one obtains $2^{n}+1$ numbers all of which are of form%
\begin{equation}
x=\frac{m}{2^{n}},  \label{13}
\end{equation}%
where $m$ is an integer and $n$ is a natural number, $n>0$. Of course, the
numbers (13) are dyadic rationals which are dense in the reals $R$. Let us
recall this theorem:

\smallskip \ 

\textbf{Theorem 2}. The set of dyadic rationals is dense in the reals $R$.

\smallskip \ 

Proof:

\smallskip \ 

Assume that $a<b$, with $a$ and $b$ elements of the reals $R$. By
Archimedean property exist $n\in N$ such that

\begin{equation}
0<\frac{1}{n}<b-a,  \label{14}
\end{equation}%
which implies%
\begin{equation*}
0<\frac{1}{2^{n}}<\frac{1}{n}<b-a.
\end{equation*}%
Thus, one has

\begin{equation}
1<2^{n}b-2^{n}a.  \label{15}
\end{equation}%
As the distance between $2^{n}b$ and $2^{n}a$ is grater than $1$, there is
an integer $m$ such that

\begin{equation}
2^{n}a<m<2^{n}b  \label{16}
\end{equation}
and therefore

\begin{equation}
a<\frac{m}{2^{n}}<b.  \label{17}
\end{equation}%
So, the set of dyadic rationals are dense in $R$.

The sum and product of surreal numbers are defined as

\begin{equation}
x+y=\{X_{L}+y,x+Y_{L}\mid X_{R}+y,x+Y_{R}\}  \label{18}
\end{equation}%
and

\begin{equation}
\begin{array}{c}
xy=\{X_{L}y+xY_{L}-X_{L}Y_{L},X_{R}y+xY_{R}-X_{R}Y_{R}\mid X_{L}y+xY_{R} \\ 
\\ 
-X_{L}Y_{R},X_{R}y+xY_{L}-X_{R}Y_{L}\}.%
\end{array}
\label{19}
\end{equation}%
The importance of (18) and (19) is that allow us to prove that the surreal
number structure is algebraically a closed field. Moreover, through (18) and
(19) it is also possible to show that the real numbers $R$ are contained in
the surreals $\mathcal{S}$ (see Ref. [1] for details).

\bigskip \ 

\noindent \textbf{3. Gonshor formalism}

\smallskip \ 

In 1986, Gonshor [3] introduced a different but equivalent definition of
surreal numbers.

\smallskip \ 

\textbf{Definition 1}. A surreal number is a function $\mu $ from initial
segment of the ordinals into the set $\{+,-\}$.

\smallskip \ 

For instance, if $\mu $ is the function so that $\mu (1)=+$, $\mu (2)=-$, $%
\mu (3)=-$, $\mu (4)=+$ then $\mu $ is the surreal number $(++-+)$. In the
Gonshor approach the expressions (3)-(5) becomes: $1$-day

\begin{equation}
-1=(-),\text{ \  \  \  \  \  \  \  \  \  \  \ }(+)=+1,  \label{20}
\end{equation}%
in the $2$-day

\begin{equation}
-2=(--),\text{ \ }-\frac{1}{2}=(-+),\text{\  \  \ }(+-)=+\frac{1}{2},\text{\  \
\  \  \  \ }(++)=+2,  \label{21}
\end{equation}%
and $3$-day

\begin{equation}
\begin{array}{c}
-3=(---),\text{ \ }-\frac{3}{2}=(--+),\text{\  \ }-\frac{3}{4}=(-+-),\text{\ }%
-\frac{1}{4}=(-++) \\ 
\\ 
(+--)=+\frac{1}{4},\text{ \ }(+-+)=+\frac{3}{4},\text{ \  \ }(++-)=+\frac{3}{2%
},\text{\  \  \  \  \  \ }(+++)=+3,%
\end{array}
\label{22}
\end{equation}%
respectively. Moreover, in Gonshor approach one finds the different numbers
through the formula

\begin{equation}
n\mid a\mid +\frac{\mid b\mid }{2}+\sum \limits_{i=i}^{q}\frac{\mid
c_{i}\mid }{2^{i+1}},  \label{23}
\end{equation}%
where $a,b,c_{1},...,c_{q}\in \{+,-\}$ and $a\neq b$. Furthermore, one has $%
\mid +\mid =+$ and $\mid -\mid =-$. As in the case of Conway definition
through (23) one gets the dyadic rationals. Observe that the values in (20),
(21) and (22) are in agreement with (23). Just for clarity, let us consider
the additional example: 
\begin{equation}
(++-+-+)=2-\frac{1}{2}+\frac{1}{4}-\frac{1}{8}+\frac{1}{16}=\frac{27}{16}.
\label{24}
\end{equation}

By the defining the order $x<y$ if $x(\alpha )<y(\alpha )$, where $\alpha $
is the first place where $x$ and $y$ differ and the convention $-<0<+$, it
is possible to show that the Conway and Gonshor definitions of surreal
numbers are equivalent (see Ref. [3] for details).

\bigskip \ 

\noindent \textbf{4. Surreal numbers and tensors}

\smallskip \ 

Let us introduce a $p$-tensor [4],

\begin{equation}
t_{\mu _{1}\mu _{2}...\mu _{p}},  \label{25}
\end{equation}%
where the indices $\mu _{1},\mu _{2},...,\mu _{p}$ run from $1$ to $2$. Of
course $p$ indicates the rank of $t_{\mu _{1}\mu _{2}...\mu _{p}}$. In
tensorial analysis, (25) is a familiar object. One arrives to a link with
surreal numbers by making the identification $1\rightarrow +$ and $%
2\rightarrow -$. For instance, the tensor $t_{1121}$ in the Gonshor notation
becomes

\begin{equation}
t_{1121}\rightarrow t_{++-+}\rightarrow (++-+).  \label{26}
\end{equation}

In terms of $t_{\mu _{1}\mu _{2}...\mu _{p}}$, the expressions (18), (19)
and (20) read

\begin{equation}
-1=t_{2},\text{ \  \  \  \  \  \  \  \  \  \ }t_{1}=+1,  \label{27}
\end{equation}%
in the $2$-day

\begin{equation}
-2=t_{22,}\text{ \  \ }-\frac{1}{2}=t_{21},\text{\  \  \ }t_{12}=\frac{1}{2},%
\text{\  \  \ }t_{11}=2,  \label{28}
\end{equation}%
and $3$-day

\begin{equation}
\begin{array}{c}
-3=t_{111},\text{ \ }-\frac{3}{2}=t_{112},\text{\  \ }-\frac{3}{4}=t_{121},%
\text{\ }-\frac{1}{4}=t_{122},\text{\ } \\ 
\\ 
t_{211}=+\frac{1}{4},\text{ \ }t_{212}=+\frac{3}{4},\text{ \ }t_{221}=+\frac{%
3}{2},\text{\  \  \ }t_{222}=+3,%
\end{array}
\label{29}
\end{equation}%
\smallskip \ respectively.

Formally, one note that there is a duality between positive and negative
labels in surreal numbers. In fact, one can prove that this is general for
any $n$-day. This could be anticipated because according to Conway
definition (1) a surreal number can be written in terms of the dual pair
left and right sets $X_{L}$ and $X_{R}$. Further, the concept of duality it
is even clearer in the Gonshor definition of surreal numbers since in such a
case one has a functions $\mu $ with codominio in the dual set $\{+,-\}$. In
terms of the tensor $t_{\mu _{1}\mu _{2}...\mu _{p}}$ in (25) such a duality
can be written in the form

\begin{equation}
t_{\mu _{1}\mu _{2}...\mu _{p}}+(-1)^{q}\varepsilon _{\mu _{1}\nu
_{1}}\varepsilon _{\mu _{2}\nu _{2}}...\varepsilon _{\mu _{p}\nu _{p}}t^{\nu
_{1}\nu _{2}...\nu _{p}}=0,  \label{30}
\end{equation}%
where

\begin{equation}
\varepsilon _{\mu \nu }=\left( 
\begin{array}{cc}
0 & 1 \\ 
-1 & 0%
\end{array}%
\right) .  \label{31}
\end{equation}

It is interesting to observe that the $2$-day corresponds to

\begin{equation}
t_{\mu _{1}\mu _{2}}=2\left( 
\begin{array}{cc}
1 & 0 \\ 
0 & -1%
\end{array}%
\right) +\frac{1}{2}\left( 
\begin{array}{cc}
0 & 1 \\ 
-1 & 0%
\end{array}%
\right) .  \label{32}
\end{equation}%
If one introduces the notation

\begin{equation}
\eta _{\mu \nu }=\left( 
\begin{array}{cc}
1 & 0 \\ 
0 & -1%
\end{array}%
\right) ,  \label{33}
\end{equation}%
one discovers that (32) can be written as

\begin{equation}
t_{\mu \nu }=2\eta _{\mu \nu }+\frac{1}{2}\varepsilon _{\mu \nu }.
\label{34}
\end{equation}

It is worth mentioning that, in general any $2\times 2$-matrix $\Omega _{\mu
\nu }$ can be written as

\begin{equation}
\Omega _{\mu \nu }=x\delta _{\mu \nu }+y\varepsilon _{\mu \nu }+r\eta _{\mu
\nu }+s\lambda _{\mu \nu }.  \label{35}
\end{equation}%
Here, one has

\begin{equation}
\delta _{\mu \nu }\equiv \left( 
\begin{array}{cc}
1 & 0 \\ 
0 & 1%
\end{array}%
\right)  \label{36}
\end{equation}%
and

\begin{equation}
\lambda _{\mu \nu }\equiv \left( 
\begin{array}{cc}
0 & 1 \\ 
1 & 0%
\end{array}%
\right) .  \label{37}
\end{equation}%
The set of matrices (31), (33), (36) and (37) determine a basis for any $%
2\times 2$-matrix belonging to the set of $2\times 2$-matrices which we
denote by $M(2,R)$.

It is interesting that by setting $r=0$ and $s=0$ in (4) one gets the
complex structure $\Omega _{\mu \nu }\longrightarrow z_{\mu \nu }$, namely

\begin{equation}
z_{\mu \nu }=x\delta _{\mu \nu }+y\varepsilon _{\mu \nu }.  \label{38}
\end{equation}%
In fact, in the typical notation of a complex number (38) becomes $z=x+iy$.
Observe also that when $\det \Omega \neq 0$ one obtains the group $GL(2,R)$
from $M(2,R)$. If one further requires that $\det \Omega =1$, then one gets
the elements of the subgroup $SL(2,R)$. It is worth mentioning that the
fundamental matrices $\delta _{\mu \nu },\eta _{\mu \nu },\lambda _{\mu \nu
} $ and $\varepsilon _{\mu \nu }$ given in (31), (33), (36) and (37) not
only form a basis for $M(2,R)$ but also determine a basis for the Clifford
algebras $C(2,0)$ and $C(1,1)$. In fact, one has the isomorphisms $%
M(2,R)\sim C(2,0)\sim C(1,1)$. There exist a theorem that establishes that
all the others higher dimensional algebras of any signature $C(a,b)$ can be
constructed from the building blocks $C(2,0),$ $C(1,1)$ and $C(0,2)$ (see
Ref. [5] and references therein). So a connection of these developments with
surreal numbers seems to be a promising scenario.

\smallskip \ 

\noindent \textbf{5. Surreal numbers and matroids}

\smallskip \ 

For a definition of a non-oriented matroid see Ref. [6] and for oriented
matroid see Ref. [7] (see also Refs. [8]-[12] and references therein). Here,
we shall focus in some particular cases of oriented matroids. First, assume
that $\chi ^{\mu \nu }$ satisfies the Grassmann-Pl\"{u}cker relation

\begin{equation}
\chi ^{\mu \lbrack \nu }\chi ^{\alpha \beta ]}=0.  \label{39}
\end{equation}%
Here, the bracket $[\nu \alpha \beta ]$ means completely antisymmetric. In
this case, the ground set of a $2$-rank oriented matroid $M=(E,\chi ^{\mu
\nu })$ is 
\begin{equation}
E=\{ \mathbf{1,2,3,4}\},  \label{40}
\end{equation}%
and the alternating map becomes%
\begin{equation}
\chi ^{\mu \nu }\rightarrow \{-1,0,1\}.  \label{41}
\end{equation}%
The $\chi ^{\mu \nu }$ function can be identified with a $2$-rank chirotope.
The collection of bases for this oriented matroid is

\begin{equation}
\mathcal{B}=\{ \mathbf{\{1,2\},\{1,3\},\{1,4\},\{2,3\},\{2,4\},\{3,4\}}\},
\label{42}
\end{equation}%
which can be obtained by just given values to the indices $\mu $ and $\nu $
in $\chi ^{\mu \nu }$. Actually, the pair $(E,\mathcal{B})$ determines a $2$%
-rank uniform non-oriented ordinary matroid.

Let us consider the underlying ground bitset (from bit and set) [13]-[14]%
\begin{equation}
\mathcal{E}=\{1,2\}  \label{43}
\end{equation}%
and the pre-ground set

\begin{equation}
E_{0}=\{(1,1),(1,2),(2,1),(2,2)\}.  \label{44}
\end{equation}%
One finds a relation between $E_{0}$ and $E$ by comparing (40) and (44). In
fact, one has

\begin{equation}
\begin{array}{cc}
(1,1)\leftrightarrow \mathbf{1}, & (1,2)\leftrightarrow \mathbf{2}, \\ 
&  \\ 
(2,1)\leftrightarrow \mathbf{3}, & (2,2)\leftrightarrow \mathbf{4}.%
\end{array}
\label{45}
\end{equation}%
This can be understood considering that (45) is equivalence relation by
making the identification of indices $\{a,b\} \leftrightarrow \mu $,..,etc.
Observe that considering this identifications the family of bases (42)
becomes

\begin{equation}
\begin{array}{c}
\mathcal{B}_{0}=\{ \{(1,1),(1,2)\},\{(1,1),(2,1)\},\{(1,1),(2,2)\}, \\ 
\\ 
\{(1,2),(2,1)\},\{(1,2),(2,2)\},\{(2,1),(2,2)\} \}.%
\end{array}
\label{46}
\end{equation}%
It turns out that the chiritope $\chi ^{\mu \nu }$ can be associated with a $%
2$-qubit system. So the pair $(E_{0},B_{0})$ can be identified with a
qubitoid (a combination of qubit and matroid).

The procedure can be generalized to higher dimensions. For instance,
consider the pre-ground set

\begin{equation}
\begin{array}{cc}
E_{0}= & \{(1,1,1,1),(1,1,1,2),(1,1,2,1),(1,1,2,2), \\ 
&  \\ 
& (1,2,1,1),(1,2,1,2),(1,2,2,1),(1,2,2,2) \\ 
&  \\ 
& (2,1,1,1),(2,1,1,2),(2,1,2,1),(2,1,2,2) \\ 
&  \\ 
& (2,2,1,1),(2,2,1,2),(2,2,2,1),(2,2,2,2)\}.%
\end{array}
\label{47}
\end{equation}%
It is not difficult to see that by making the identifications

\begin{equation}
\begin{array}{ccc}
(1,1,1,1)\leftrightarrow \mathbf{1} & (1,1,1,2)\leftrightarrow \mathbf{2} & 
(1,1,2,1)\leftrightarrow \mathbf{3} \\ 
&  &  \\ 
(1,1,2,2)\leftrightarrow \mathbf{4} & (1,2,1,1)\leftrightarrow \mathbf{5} & 
(1,2,1,2)\leftrightarrow \mathbf{6} \\ 
&  &  \\ 
(1,2,2,1)\leftrightarrow \mathbf{7} & (1,2,2,2)\leftrightarrow \mathbf{8} & 
(2,1,1,1)\leftrightarrow \mathbf{9} \\ 
&  &  \\ 
(2,1,1,2)\leftrightarrow \mathbf{10} & (2,1,2,1)\leftrightarrow \mathbf{11}
& (2,1,2,2)\leftrightarrow \mathbf{12} \\ 
&  &  \\ 
(2,2,1,1)\leftrightarrow \mathbf{13} & (2,2,1,2)\leftrightarrow \mathbf{14}
& (2,2,2,1)\leftrightarrow \mathbf{15} \\ 
&  &  \\ 
(2,2,2,2)\leftrightarrow \mathbf{16,} &  & 
\end{array}
\label{48}
\end{equation}%
one obtains a relation between the pre-ground set $E_{0}$ given in (47) and
the ground set 
\begin{equation}
E=\{ \mathbf{1,2,...,15,16}\}.  \label{49}
\end{equation}%
This can be again understood by considering that (49) is equivalent to make
the identification of indices $(a,b,c,d)\leftrightarrow \mu ,...etc$. It
turns out that considering these relations one finds that the collection of
bases $\mathcal{B}$ contains $\left( 
\begin{array}{c}
16 \\ 
2%
\end{array}%
\right) =120$ two-element subset of the $16$-element set $E$, given in (49).
This $2$-element subset can be obtained by considering a lexicographic order
of all $120$ two-subsets of $\{ \mathbf{1,2,...,15,16}\}$. One finds that
the first terms of $\mathcal{B}_{0}$ look like

\begin{equation}
\begin{array}{c}
\mathcal{B}_{0}=\{ \{(1,1,1,1),\{1,1,1,2)\},\{(1,1,1,1),\{1,1,2,1)\}, \\ 
\\ 
\{(1,1,1,1),\{1,1,2,2)\},\{(1,1,1,1),\{1,2,1,1)\},\{(1,1,1,1),\{1,2,1,2)%
\},...\}.%
\end{array}
\label{50}
\end{equation}%
(See Refs. [13] and [14] for details.)

The method, of course, can be extended to $2^{2n+1}$-dimensions, $%
n=0,1,2,...etc$ and can be connected to $N$-qubit system. However, it is
worth mentioning that the complete classification of $N$-qubit systems is a
difficult, or perhaps an impossible task. In reference [15] an interesting
development for characterizing a subclass of $N$-qubit entanglement has been
considered. An attractive aspect of this construction is that the $N$-qubit
entanglement can be understood in geometric terms. The idea is based on the
bipartite partitions of the Hilbert space in the form $C^{2^{N}}=C^{L}%
\otimes C^{l}$, with $L=2^{N-n}$ and $l=2^{n}$. Such a partition allows a
geometric interpretation in terms of the complex Grassmannian variety $%
Gr(L,l)$ of $l$-planes in $C^{L}$ via the Pl\"{u}cker embedding. In this
case, the Plucker coordinates of the Grassmannians are natural invariants of
the theory.

There are a number of ways in which one can connect matroids with surreal
numbers. First, one may think in the bitset given in (43) in the Gonshor form%
\begin{equation}
\mathcal{E}=\{1,2\} \rightarrow \{-,+\}.  \label{51}
\end{equation}%
Second, the numbers of any the ground set in matroid theory

\begin{equation}
E=\{ \mathbf{1,2,3,...}\},  \label{52}
\end{equation}%
can be written in terms of the surreal numbers as%
\begin{equation}
\{ \{0,\emptyset \},\{1,\emptyset \},\{2,\emptyset \},...\}.  \label{53}
\end{equation}%
In this context the basis set $\mathcal{B}$ will be also written in terms of
the surreal numbers. Third, another possibility is also to identify the
chirotope map $\chi \rightarrow \{-1,0,1\}$ in terms of the surreal numbers $%
\chi \rightarrow \{ \{ \emptyset ,0\},\{ \emptyset ,\emptyset
\},\{0,\emptyset \} \}$.

Of course, it will interesting to fully develop these possible links between
matroids and surreal numbers. But even at these stage one note that the key
concept in both matroid theory and surreal numbers theory is duality. This
is because in matroid theory it is known that in matroid theory there is a
key theorem that every matroid $\mathcal{M}$ has a dual $\mathcal{M}^{\ast }$%
, while in surreal number theory duality is every where. In a sense this is
because a surreal numbers $x=\{X_{L},X_{R}\}$ is defined in terms of two
dual sets $X_{L}$ and $X_{R}$. So one wonders whether in surreal number
theory exist a theorem establishing that for every surreal number set $%
\mathcal{S}$ there exist a dual surreal number set $\mathcal{S}^{\ast }$.

\smallskip \ 

\noindent \textbf{6. Various mathematical and physical possible applications}

\smallskip \ 

In this section we shall describe an additional number of possible
applications of surreal numbers in mathematics and physics. Although such a
description will be brief the main idea is to stimulate further research in
the area. One may think that our proposals are in a sense for experts in the
topic but in fact the main intention is to call the attention of
mathematicians and physicist telling them look here are a number of subjects
in which you have the opportunity to participate.

\bigskip \ 

\textit{I. Applications in mathematics:}

\smallskip \ 

\textit{(a) Division algebras}

\smallskip \ 

There is a celebrated Hurwitz theorem:

\smallskip \ 

\textbf{Theorem (Hurwitz, 1898): }\textit{Every normed algebra over the
reals with an identity is isomorphic to one of following four algebras: the
real numbers, the complex numbers, the quaternions, and the Cayley
(octonion) numbers.}

\smallskip \ 

Moreover, the Hurwitz theorem is closely related with the parallelizable
spheres $S^{1},S^{3}$ and $S^{7}$ [16] and the remarkable theorem that only
exist division algebras in $1$, $2$, \ $4$ and $8$ dimensions [17]-[18]. So,
one wonders what could be the corresponding Hurwitz theorem and these
remarkable developments on division algebras if one extend the real numbers
to surreal numbers. In this context, it has been proved in Refs. [19]-[21]
that for normalized qubits the complex $1$-qubit, $2$-qubit and $3$-qubit
are deeply related to division algebras via the Hopf maps, $S^{3}\overset{%
S^{1}}{\longrightarrow }S^{2}$, $S^{7}\overset{S^{3}}{\longrightarrow }S^{4}$
and $S^{15}\overset{S^{7}}{\longrightarrow }S^{8}$, respectively. It seems
that there does not exist a Hopf map for higher $N$-qubit states. So, from
the perspective of Hopf maps, and therefore of division algebras, one
arrives to the conclusion that $1$-qubit, $2$-qubit and $3$-qubit are more
special than higher dimensional qubits (see Refs. [22]-[23] for details).
Again one wonders whether surreal numbers can contribute in this qubits
theory framework.

\smallskip \ 

\textit{II. Applications in physics:}

\smallskip \ 

\textit{(a) Supersymmetry}: For finite sets $X_{L}$ and $X_{R}$, one of the
key tools in surreal numbers are integers $n$ and dyadic rationals $d=\frac{m%
}{2^{k}}$. For $n=0,1$ and $2$ and $d=\frac{1}{2},\frac{3}{2}$ one recalls
the spin structure of supersymmetry. So one wonders if for instance spin $%
\frac{1}{4}$ may be a prediction of surreal number theory. Remarkable, this
spin has been proposed in $N=1$ supersymmetry in connection with anyons (
see Refs. [24]-[25] and references therein). Thus, for finite sets $X_{L}$
and $X_{R}$, surreal numbers $x=\{X_{L}$,$X_{R}\}$ and in the Gonshor
approach, one finds that bosons can be identified with $s$ integer spin and
fermions with dyadic rational with spin%
\begin{equation*}
n\mid a\mid +\frac{\mid b\mid }{2}+\sum \limits_{i=1}^{q}\frac{\mid
c_{i}\mid }{2^{i+1}},
\end{equation*}%
given in (23). One can even think in this expression as the eigenvalues of a
ket $\mid n,s_{1},s_{2},s_{3,}...>$. Here we made the associations%
\begin{equation}
s_{1}\rightarrow \frac{\mid b\mid }{2},s_{2}\rightarrow \frac{\mid c_{1}\mid 
}{4},s_{3}\rightarrow \frac{\mid c_{2}\mid }{8}...  \label{54}
\end{equation}%
and so on. Thus, in this framework, it seems the whole structure of surreal
numbers can be identified with a kind of supersymmetric approach.

\smallskip \ 

\textit{(c)} \textit{Black-holes }

\smallskip \ 

Consider the Schwarzschild metric [26]

\begin{equation}
ds^{2}=-(1-\frac{2GM}{c^{2}r})dt^{2}+\frac{dr^{2}}{(1-\frac{2GM}{c^{2}r})}%
+r^{2}(d\theta ^{2}+sen^{2}\theta d\phi ^{2}),  \label{55}
\end{equation}%
where $M$ is the source mass, $G$ is the Newton gravitational constant and $%
c $ is the light velocity. There are a number of observations that one can
make about (55). First, notice that in this expression all quantities are
real numbers. Second there are two type of singularities, namely in $r=r_{s}=%
\frac{2GM}{c^{2}}$ and $r=0$. It is known that using Kruskal coordinates it
is possible to show that the singularity at $r=r_{s}$ is simply a coordinate
singularity. However the singularity at $r=0$ is a true physical singularity
of spacetime. First of all, in this context, when one referes about
singularity in terms of real numbers one means that in the limit $%
r\rightarrow 0$ one obtains the expression $\frac{2GM}{c^{2}r}\rightarrow
\infty $ (see Ref. [26] and references therein for details).

From the point of view of surreal numbers theory the singularity $\frac{2GM}{%
c^{2}r}\rightarrow \infty $, when $r\rightarrow 0$, is not a real problem
because in such a mathematical theory all kind of infinite large and
infinite small are present. So by a assuming that all quantities in the line
element given in (55) is written in terms of surreal numbers

\begin{equation}
d\mathcal{S}^{2}=-(1-\frac{2G\mathcal{M}}{c^{2}\mathcal{R}})d\mathcal{T}^{2}+%
\frac{d\mathcal{R}^{2}}{(1-\frac{2G\mathcal{M}}{c^{2}\mathcal{R}})}+\mathcal{%
R}^{2}(d\mathcal{\theta }^{2}+sen^{2}\mathcal{\theta }d\mathcal{\phi }^{2}),
\label{56}
\end{equation}%
the problem of singularities in black-hole physics no longer exist!

\smallskip \ 

\textit{(d) Cosmology}

\smallskip \ 

In the Friedmann cosmological equation [26]

\begin{equation}
\frac{1}{a^{2}}\frac{da(t)}{dt}+\frac{k}{a^{2}}-\frac{8\pi \rho _{m}}{3}-%
\frac{8\pi \rho _{r}}{3}=0,  \label{57}
\end{equation}%
one assumes that the matter density is given by

\begin{equation}
\rho _{m}=\frac{\rho _{0m}}{a^{3}},  \label{58}
\end{equation}%
while the radiation energy density is

\begin{equation}
\rho _{r}=\frac{\rho _{0r}}{a^{4}},  \label{59}
\end{equation}%
where $k,\rho _{0m}$ and $\rho _{0r}$ are constants. So, even if one does
not consider the solution of (57) the expressions (58) and (59) tell us that
there is a `big-bang' singularity at $a\rightarrow 0$. In fact, when $%
a\rightarrow 0$ one has%
\begin{equation}
\rho _{m}\rightarrow \infty  \label{60}
\end{equation}%
and%
\begin{equation}
\rho _{r}\rightarrow \infty .  \label{61}
\end{equation}%
Just as in the case of black-holes these singularities are related to the
fact that one is considering real numbers structure in the length scale $a$
as well as in the time evolution parameter $t$. Again, one wonders what
formalism one may obtain by replacing $\ a$ by some kind of surreal length
scale $\mathcal{A}$ and the time parameter $t$ by a surreal time parameter $%
\mathcal{T}$. Of course, this in turn will imply that the whole
gravitational theory must be modified with surreal numbers structure.

Another possibility is to identify the whole evolution of the surreal
numbers structure with a cosmological model in the sense that in $0$-day one
has the scalar field particle of $0$-spin (the Higgs field?), in the $1$-day
one has $(-1)$-spin and $1$-spin (the photon?) and the $2$-day one obtains
the $(-2)$-spin, $(-\frac{1}{2})$-spin, $\frac{1}{2}$-spin $2$-spin
(graviton and fermion?) and so on. Following this idea one may even identify
the $0$-day and $0$-spin with the big bang and since $0=\{ \emptyset
,\emptyset \}$ one can say that everything in our universe started with
vacuum state $\emptyset $.

\smallskip \ 

\textit{(e) Fractals}

\smallskip \ 

It is known that fractals and dyadic fractions are deeply related. Much of
this relationship can be explained by infinite binary tree which can be
viewed as a certain subset of the modular group $PSL(2,Z)$ (the general
linear group of $2$ by $2$ matrices over the integers). The subset is
essentially the dyadic grupoid or dyadic monoid. This in turn provides the
natural setting for the symmetry and self-similarity of many fractals.
Moreover, it is also that these groups and the rational numbers \ can be
connected with dyadic subsets [27].

\smallskip \ 

\noindent \  \textbf{7. Final remarks}

\smallskip \ 

Due to the fact that duality is the underlying concept in both surreal
numbers and matroid theory, we believe that it is a matter of time that
these two mathematical scenarios are considered as important tools in
physics and in particular in high energy physics and gravity.

From the serious difficulties with infinities in black-hole physics and
cosmology as well as in higher energy physics it seems to us that surreal
numbers theory offers a new view for a solution. Instead of thinking that
the infinities are the enemies in quantum and classical physical theory
incorporate them in a natural way as surreal numbers framework suggest.

It turns out that surreal numbers can be understood as a particular case of
games [2] (see also Ref. [31]) which is a fascinating mathematical theory.
In fact, games can be added and substracted forming an abelian group and a
sub-group of games is identified with surreal numbers which can also be
multiplied and form a field. As we mentioned before this field contains the
real numbers among many other numbers structures. The key additional
condition for reducing a game $x=\{x_{L},...,x_{R},...\}$ to a surreal
number is that $x_{L}$ and $x_{R}$ are surreal numbers and satisfy $%
x_{L}<x_{R}$. So, one wonders whether game theory may lead to even more
interesting applications that those presented in this work.

Finally, we believe that it is just a matter of time for the recognition of
the surreal numbers structure as one of the key mathematical tools in
superstring theory [28]-[30]. This is because although the problems of some
infinities are solved there remain always additional problems with the
emergency of new infinities. This phenomena may be traced back to the fact
that the action in superstring theory is written in terms of real functions
(target space-time coordinates) rather that surreal functions.

\bigskip \ 

\noindent \textbf{Acknowledgments: }I would like to thank to P. A. Nieto, C.
Garc\'{\i}a-Quintero and A. Meza for helpful comments. This work was
partially supported by PROFAPI/2013.

\smallskip \

\end{document}